\documentstyle[preprint,eqsecnum,aps,epsf]{revtex}
\begin{document}
\tolerance=10000
\hfuzz= 5 pt
\baselineskip=24pt
\draft
\preprint{IFUM 533/FT, June 1996}

\title{Heavy Quarkonia: Wilson Area Law, Stochastic Vacuum Model and
Dual QCD}
\author {N. Brambilla}
\address{Dipartimento di Fisica dell'Universit\`{a}, Milano, \\
INFN, Sezione di Milano, Via Celoria 16, 20133 Milano, Italy}
\author {A. Vairo}
\address {Dipartimento di Fisica, Universit\`a di Bologna, \\
 Via Irnerio 46, 40126 Bologna, Italy}
\maketitle

\begin{abstract}
\baselineskip=20pt
The $Q \bar{Q} $ semirelativistic interaction in QCD can be simply expressed 
in terms of the Wilson loop and its functional derivatives. In this approach 
we present the $Q \bar{Q}$ potential up to order $1/m^2$ using the 
expressions for the Wilson loop given by the Wilson Minimal Area Law (MAL), 
the Stochastic Vacuum Model (SVM) and Dual QCD (DQCD). We confirm 
the original results given in the  different frameworks and obtain new 
contributions. In particular we calculate up to order $1/m^2$ the complete 
velocity dependent potential in the SVM. This allows us to show that 
the MAL model is entirely contained in the SVM. We compare and discuss 
also the SVM and the DQCD potentials. It turns out that in these two very 
different models the spin-orbit potentials show up the same leading 
non-perturbative contributions and $1/r$ corrections in the long-range limit.
\end{abstract}

\pacs{PACS numbers:12.38.Aw, 12.38.Lg, 12.39.Pn}

\section{INTRODUCTION}

Since the pioneering paper of Wilson \cite{Wilson} a real 
breakthrough opened in the treatment of quark states and in this framework 
a lot of work was devoted to the study of the heavy $Q \bar{Q}$. 
The challenge was understanding low energy QCD dynamics and hence confinement. 
The main characteristics of the heavy meson and baryon spectrum are 
simple and cleanly connected to expectation value of the $Q \bar{Q}$ 
and $3 Q$ potentials. The size of the $b$ and $c$ systems extends over  
distances where confinement already plays a relevant role (only toponium 
can be described purely in terms of one gluon exchange plus 
higher order perturbative corrections \cite{Kummer} but, as well-known, 
we cannot access its spectrum); moreover, due to the mean value 
of the quark velocities, the leading relativistic corrections 
can be appreciated and usefully tested on the data. 
Furthermore, a good understanding of the heavy quark semirelativistic 
interaction is the first step towards relativistic generalization. 

At the static level, the linear confining $Q \bar{Q}$ 
interaction, corresponding to a constant energy density 
(the string tension $\sigma$) localized in a flux tube between 
the quarks, emerges in lattice formulation of QCD and is 
contained in all the existing confining models, e.g., Wilson area law, 
flux tube model and all kind of dielectric and dual models. 
This corresponds also to the static limit of 
the Buchm\"uller's picture \cite{Buch}  of a 
rotating quark-antiquark state connected  by a purely chromoelectric 
tube with a pure transverse velocity and with chromomagnetic field 
vanishing in  the comoving system of the tube. 
In this picture it follows simply  that the non-perturbative 
spin-interaction is given only by the Thomas precession term. 

The spin-dependent relativistic corrections were calculated first 
by Eichten, Feinberg \cite{eichten} and Gromes \cite{Gromes} 
as a correction to the static limit (Wilson--Brown--Weisberger area 
law result). The potential is expressed in terms of average of electric 
and magnetic fields that can also be calculated on the lattice. 
The Eichten--Feinberg--Gromes results, at least in the long range 
behaviour, have been reproduced on the lattice \cite{lattice,lattice3} (for a 
detailed discussion see Sec. 6). Recently the spin-dependent potential was 
also studied in the context of the Heavy Quark Effective Theory \cite{HQET}.

In the literature relativistic generalizations of these results were 
attempted in a Bethe--Salpeter context by constructing a Bethe--Salpeter 
kernel which give back static and spin-dependent potentials. 
Using a simple convolution kernel (i.e. depending only on the 
momentum transfer $Q$), this amounts to considering a Lorentz 
scalar proportional to $1/Q^4$. The velocity dependent  
relativistic corrections were also obtained but they are 
strongly dependent on the type of ``instantaneous" approximation 
chosen to define the potential and on the gauge. 
These non-perturbative velocity dependent corrections 
destroy the agreement with the data \cite{Gupta,Durand,Lagae} 
and give origin to the puzzle of how reconciling the spin-structure 
(i.e. the Lorentz nature of the kernel) with the velocity corrections
in one Bethe--Salpeter kernel. In this paper we will not deal with 
this problem starting directly from the $1/m^2$ expansion of the 
potential without any relativistic assumption. 
However a first step in its resolution seems to be the 
correct inclusion of the low energy dynamics also in the 
spin-independent $1/m^2$ corrections. Moreover from the knowledge 
of these and the spin dependent  corrections we will obtain some  
important insights on the nature of the kernel. 

Recently a method to obtain the complete $1/m^2$ quark--antiquark 
(and 3 quarks) potential, based on the path integral representation 
of the Pauli--type quark propagator, was given 
in \cite{BCP} (see also \cite{BMP,NC} and \cite{Sinp}). This formulation 
is gauge invariant. The potential is obtained as a function of a generalized 
Wilson loop (i.e. any kind of trajectory for the quark and the antiquark 
can appear) and its functional derivatives. These are all measurable on the 
lattice.  In short it was obtained a constituent quark semirelativistic 
interaction with coefficients determined by the non linear gluodynamics. 
This is the ideal framework in which to formulate hypothesis 
on the Wilson loop behaviour (and so on the confinement 
mechanism) to be checked on the lattice and on the experimental data. 

First, to evaluate the non-perturbative behaviour of the Wilson 
loop, a modified minimal area law (MAL) was used (see Sec. 3). 
This reproduces the Eichten--Feinberg--Gromes results  
\cite{eichten,Gromes} and gives a velocity dependent potential proportional 
to the flux tube angular momentum squared, so that, by including velocity 
dependent corrections, a ``string model'' emerges (see \cite{Lagae,Dubin}). 
Also the velocity dependent potentials seem to agree with recent 
available lattice data \cite{Bali}. 

However, the MAL represents an extreme approximation 
that gives the correct result for very large interquark distances 
and does not give insight into open problems such as the relation 
between the non-perturbative structure of spin and velocity 
corrections. For these reasons we have taken into 
account two models of confinement, the stochastic vacuum 
model (SVM) and Dual QCD (DQCD) which both give an expression 
for the whole behaviour of the Wilson loop and contain 
the area law in the long distances limit.
It is interesting to realize that both models reproduce essentially 
the perturbative plus MAL results respectively in the limit 
of short and long distances but produce also subleading  
corrections.  These allows us to understand better the physical 
picture. For example in the case of the non-perturbative spin-orbit 
interaction it turns out that the magnetic term cancels 
in the area law limit (zero magnetic field in the comoving framework) but 
presents $1/r$ suppressed corrections in the other two models. 

A careful comparison between the SVM and DQCD corrections 
and an investigation of the approximations in which they coincide seem 
to be of great importance to the aim of understanding the low energy 
gluodynamics contained in the Wilson loop. 

The plan of the paper is the following one. In Sec. 2 we briefly 
sum up the definition of the semirelativistic potential and the notations. 
In Sec. 3 we collect the results obtained in the MAL model. 
In Sec. 4 we briefly  present the SVM and use it to evaluate the 
potential in the context of Sec. 2. In particular we obtain also the SVM 
velocity dependent potential which is new. We show that it satisfies 
important identities and we give the short and long-range limits. 
In Sec. 5 we introduce the DQCD potential and discuss the long-range limit. 
In Sec. 6 we discuss our results in connection with the up to now available 
lattice data and draw some conclusions.
   
\section{THE QUARK-ANTIQUARK POTENTIAL}

In \cite{BCP} a Foldy--Wouthuysen transformation  
on the quark-antiquark Green's function was done and the result was 
written as a Feynman path integral over particle and 
anti-particle coordinates and momenta of a Lagrangian depending only upon 
the spin, coordinates, and momenta of the quark and antiquark.  
Separating off the kinetic terms from this Lagrangian it was possible 
to identify the heavy quark potential $V_{{\rm Q} \bar {\rm Q}}$ 
(closed loops of light quark pairs and annihilation contributions were 
not included):
\begin{eqnarray}
\int_{t_{\rm i}}^{t_{\rm f}} dt V_{{\rm Q} \bar{{\rm Q}}} &=& 
i \log \langle W(\Gamma) \rangle 
- \sum_{j=1}^2 \frac{g}{m_j} \int_{{\Gamma}_j}dx^{\mu} 
\left( S_j^l \, 
\langle\!\langle \hat{F}_{l\mu}(x) \rangle\!\rangle  
-\frac{1}{2m_j} S_j^l \varepsilon^{lkr} p_j^k \, 
\langle\!\langle F_{\mu r}(x) \rangle\!\rangle \right.
\nonumber\\
&~& \quad\quad - \left. \frac{1}{8m_j} \, 
\langle\!\langle D^{\nu} F_{\nu\mu}(x) \rangle\!\rangle  \right)
- \frac{1}{2} \sum_{j,j^{\prime}=1}^2 \frac{ig^2}{m_jm_{j^{\prime}}}
{\rm T_s} \int_{{\Gamma}_j} dx^{\mu} \, \int_{{\Gamma}_{j^{\prime}}} 
dx^{\prime\sigma} \, S_j^l \, S_{j^{\prime}}^k
\nonumber\\
&~& \quad\quad \times \left( \, 
\langle\!\langle \hat{F}_{l \mu}(x) \hat{F}_{k \sigma}(x^{\prime})
\rangle\!\rangle  - \, 
\langle\!\langle \hat{F}_{l \mu}(x) \rangle\!\rangle
\, \langle\!\langle \hat{F}_{k \sigma}(x^{\prime}) \rangle\!\rangle  \right) 
\> ,
\label{potential}
\end{eqnarray}
where
\begin{eqnarray}
F_{\mu\nu} &=& \partial_\mu A_\nu - \partial_\nu A_\mu 
+ i g \, [A_\mu, A_\nu] \, ,
\quad \quad \quad 
\hat{F}^{\mu \nu} \equiv {1\over 2} \varepsilon^{\mu \nu \rho \sigma}  
F_{\rho\sigma} \, , 
\\
D^\nu F_{\nu\mu} &=& \partial^\nu F_{\nu\mu} + ig[A^\nu,F_{\nu\mu}] \>, 
\\
W(\Gamma) &\equiv&  
{\rm P\>} \exp \left[i g \oint_{\Gamma} dx^\mu A_\mu (x) \right] \> ,
\label{wgamma}
\end{eqnarray}
and 
\begin{eqnarray}
\langle f(A) \rangle &\equiv& 
{1\over 3}{\rm Tr \>}{\rm P \>}
{\int {\cal D} A \, e^{iS_{\rm YM} (A)} f(A) 
\over \int {\cal D} A\,  e^{iS_{\rm YM} (A)}} \>,
\\
\langle\!\langle f(A) \rangle\!\rangle &\equiv&
{\int {\cal D} A \, e^{iS_{\rm YM} (A)} 
{\rm Tr \>}{\rm P \>} \{ f(A) 
\exp \left[i g \oint_{\Gamma} dx^\mu A_\mu (x) \right] \}
\over \int {\cal D} A\,  e^{iS_{\rm YM} (A)} {\rm Tr \>}{\rm P \>}
{\exp \left[i g \oint_{\Gamma} dx^\mu A_\mu (x) \right] } } \>.
\end{eqnarray}
The closed loop $\Gamma$ is defined by the quark (anti-quark) 
trajectories ${\bf z}_1 (t)$ (${\bf z}_2 (t)$) running from ${\bf y}_1$ 
to ${\bf x}_1$ (${\bf x}_2$ to ${\bf y}_2$) as $t$ varies from the 
initial time $t_{\rm i}$ to the final time $t_{\rm f}$.
The quark (anti-quark) trajectories ${\bf z}_1(t)$  (${\bf z}_2 (t)$) 
define the world lines $\Gamma_1$ ($\Gamma_2$) running from $t_{\rm i}$ to 
$t_{\rm f}$  ($t_{\rm f}$ to $t_{\rm i}$). 
The world lines $\Gamma_1$ and $\Gamma_2$, along with 
two straight-lines at fixed time connecting ${\bf y}_1$ to ${\bf y}_2$ 
and ${\bf x}_1$ to ${\bf x}_2$, then make up the contour $\Gamma$ 
(see Fig. 1). 
\footnote
{As a consequence $\int_{\Gamma_j} dx^{\mu}
f_{\mu}(x) = (-1)^{j+1} \int_{t_{\rm i}}^{t_{\rm f}} dt ( f_0(z_j) 
- {\dot{{\bf z}}}_j \cdot {\bf f} (z_j))$, 
where $z_j=(t,{\bf z}_j(t))$. 
The factor $(-1)^{j+1}$  accounts for the fact that 
world line $\Gamma_2 $  runs from $t_{\rm f}$ to $t_{\rm i}$.
We also use the notation 
$z_j^{\prime}=(t^{\prime},{\bf z}_j(t^{\prime}))$.} 
As usual $A_\mu (x) \equiv A_\mu^a (x) \lambda_a/2$, 
${\rm Tr}$ means the trace over color indices,  ${\rm P}$
prescribes the ordering of the color matrices according 
to the direction fixed on the loop and $S_{\rm YM}(A)$ is the Yang--Mills 
action including a gauge fixing term.

As the  $1/m^2$  terms in $V_{{\rm Q} \bar {\rm Q}}$  are of 
two types, velocity dependent $V_{\rm VD}$ and spin dependent $V_{\rm SD}$, 
we can identify in the full potential three type of contributions:
\begin{equation} 
V_{{\rm Q} \bar {\rm Q}} = V_0 + V_{\rm VD} + V_{\rm SD} \>,
\end{equation}
with $V_0$ the static potential. 

The spin independent part of the potential, $V_0 + V_{\rm VD}$, is 
obtained in (\ref{potential}) from the zero order and the quadratic 
terms in the expansion of $\log \langle W(\Gamma) \rangle $ for 
small velocities ${\dot{\bf z}}_1(t) = {\bf p}_1/m_1$  and 
${\dot{\bf z}}_2(t) = {\bf p}_2/m_2$.  
In the notation of \cite{BMP,dualin} 
the terms arising from this expansion can be rearranged as:
\begin{eqnarray}
i \log \langle W(\Gamma) \rangle &=& \int_{t_{i}}^{t_{f}} dt ~
V_0 (r(t)) + V_{\rm VD}({\bf r}(t)) \, ,
\\
V_{\rm VD}({\bf r}(t)) &=&  {1\over m_1 m_2} 
\left\{ {\bf p}_1\cdot{\bf p}_2 V_{\rm b}(r) 
+ \left( {1\over 3} {\bf p}_1\cdot{\bf p}_2 - 
{{\bf p}_1\cdot {\bf r} \>~ {\bf p}_2 \cdot {\bf r} \over r^2}\right) 
V_{\rm c}(r) \right\}_{\rm Weyl} 
\nonumber \\
&+& \sum_{j=1}^2 {1\over m_j^2}
\left\{ p^2_j V_{\rm d}(r) 
+ \left( {1\over 3} p^2_j - 
{{\bf p}_j\cdot {\bf r} \>~ {\bf p}_j \cdot {\bf r} \over r^2}\right) 
V_{\rm e}(r) \right\}_{\rm Weyl} \, , 
\label{vd}
\end{eqnarray}
where ${\bf r}(t) \equiv {\bf z}_1(t) - {\bf z}_2 (t)$ and 
the symbol $\{\> ~\}_{\rm Weyl}$ stands for the Weyl ordering 
prescription among momentum and position variables \cite{BCP}.

The spin dependent potential $V_{\rm SD}$ contains for 
each quark terms analogous to those obtained by making a Foldy--Wouthuysen 
transformation on the Dirac equation in an external field 
(where $\langle\!\langle F_{\mu \nu} \rangle\!\rangle$ plays the role of the 
external field), along with an additional term $V_{\rm SS}$ having the 
structure of a spin-spin interaction.  We can then write
\begin{equation}
V_{\rm SD} = V_{\rm LS}^{\rm MAG} + V_{\rm Thomas} + V_{\rm Darwin} 
           + V_{\rm SS} \>,
\label{vsd}
\end{equation}
using a notation which indicates the physical significance of the 
individual terms (MAG denotes Magnetic). The correspondence 
between (\ref{vsd}) and (\ref{potential}) is given by 
\begin{eqnarray}
\int_{t_{\rm i}}^{t_{\rm f}} dt V_{\rm LS}^{\rm MAG} &=& 
- \sum_{j=1}^2 \frac{g}{m_j} \int_{{\Gamma}_j}dx^{\mu} 
S_j^l \, \langle\!\langle \hat{F}_{l\mu}(x) \rangle\!\rangle  
\>, \label{vmag} \\
\int_{t_{\rm i}}^{t_{\rm f}} dt V_{\rm Thomas} &=& 
\sum_{j=1}^2 \frac{g}{2 m^2_j} \int_{{\Gamma}_j}dx^{\mu} 
S_j^l \varepsilon^{lkr} p_j^k \, 
\langle\!\langle F_{\mu r}(x) \rangle\!\rangle 
\>, \label{vthomas} \\
\int_{t_{\rm i}}^{t_{\rm f}} dt V_{\rm Darwin} &=& 
\sum_{j=1}^2 \frac{g}{8 m^2_j} \int_{{\Gamma}_j}dx^{\mu} 
\langle\!\langle D^{\nu} F_{\nu\mu}(x) \rangle\!\rangle  
\>, \label{vdarwin} \\
\int_{t_{\rm i}}^{t_{\rm f}} dt V_{\rm SS} &=& 
- \frac{1}{2} \sum_{j,j^{\prime}} \frac{ig^2}{m_jm_{j^{\prime}}}
{\rm T_s} \int_{{\Gamma}_j} dx^{\mu} \, \int_{{\Gamma}_{j^{\prime}}} 
dx^{\prime\sigma} \, S_j^l \, S_{j^{\prime}}^k
\left( \, \langle\!\langle \hat{F}_{l \mu}(x) \hat{F}_{k \sigma}(x^{\prime})
\rangle\!\rangle  \right.
\nonumber\\
&~& \quad\quad\quad\quad\quad\quad
\left. - \langle\!\langle \hat{F}_{l \mu}(x) \rangle\!\rangle
\, \langle\!\langle \hat{F}_{k \sigma}(x^{\prime}) \rangle\!\rangle  \right) 
\> . \label{vss}
\end{eqnarray}
In the well-known Eichten and Feinberg notation \cite{eichten} and 
taking also into account the Darwin potential and similar 
contributions arising from the spin-spin interaction \cite{BMP,dualin}, 
the terms in $V_{\rm SD}$ can be rearranged as 
\begin{eqnarray}
V_{\rm SD} &=& 
{1\over 8} \left( {1\over m_1^2} + {1\over m_2^2} \right) 
\Delta \left[ V_0(r) +V_{\rm a}(r) \right] 
\nonumber\\
&+& \left( {1\over 2 m_1^2} {\bf L}_1 \cdot {\bf S}_1 
     - {1\over 2 m_2^2} {\bf L}_2 \cdot {\bf S}_2 \right) 
       {1\over r}  {d \over dr} \left[ V_0(r)+ 2 V_1(r) \right]
\nonumber \\
&+&
{1\over m_1 m_2}
\left( {\bf L}_1 \cdot {\bf S}_2 - {\bf L}_2 \cdot {\bf S}_1 \right) 
{1\over r} {d \over dr} V_2(r) 
+{1\over m_1 m_2}  
\left( { {\bf S}_1\cdot{\bf r} \> {\bf S}_2\cdot{\bf r}\over r^2} 
- {1\over 3} {\bf S}_1 \cdot {\bf S}_2 \right) V_3(r) 
\nonumber \\
&+& {1\over 3 m_1 m_2} 
{\bf S}_1 \cdot {\bf S}_2 \, V_4(r) \>,
\label{sd}
\end{eqnarray}
with ${\bf L}_j = {\bf r} \times {\bf p}_j$.
It is not possible to identify directly each Eichten and Feinberg 
potential with the terms contained in eq. (\ref{potential}) without 
making some assumptions on the Wilson loop. This will be the aim 
of the next sections. But some observations are just now possible.
The contributions to $\Delta (V_0 +V_{\rm a})$ 
come from $V_{\rm Darwin}$ and from $V_{\rm SS}$ with $j = j^\prime$. 
In the case $j \neq j^\prime$, $V_{\rm SS}$ contributes to 
the tensor term $V_3$ and to the spin-spin term $V_4$. 
Finally, $V_1$ receives contributions from both the magnetic 
($V_{\rm LS}^{\rm MAG}$) and the Thomas precession term 
($V_{\rm Thomas}$) while the contributions to $V_2$ come 
only from the magnetic term.

Due to the Lorentz invariance properties of the Wilson loop some exact 
relations for the potentials $V_i$ and $V_{\rm a}$, ..., $V_{\rm e}$ 
can be obtained. The first was given by Gromes \cite{Gromes} 
for the spin-related potentials
\begin{equation}
{d\over dr} \left[ V_0(r) +V_1(r)-V_2(r) \right] = 0 \> ,
\label{grom}
\end{equation}
and the other one  by Barchielli, Brambilla and Prosperi \cite{NC}
for the velocity-related potentials
\begin{eqnarray}
& & V_{\rm d}(r) +{1\over 2} V_{\rm b}(r) +{1\over 4} V_0(r) - 
{r\over 12} {d V_0(r)\over dr}=0  \> ,
\label{relvel1}\\
& &V_{\rm e}(r) +{1\over 2} V_{\rm c}(r) 
+ {r\over 4} {dV_0(r)\over dr}=0  \> .
\label{relvel2}
\end{eqnarray}
Since these relations are due to the Lorentz invariance they must be
satisfied by any good choice of the Wilson loop approximated behaviour.

Summarizing, the static and velocity dependent part of the potential
are given in terms of the expansion of the Wilson loop average 
$\langle W (\Gamma) \rangle $, while 
the spin dependent potentials are given as a sum of terms depending 
upon the quark and antiquark spins, masses and momenta with coefficients 
which are expectation values of operators computed in presence 
of a moving quark-antiquark pair.  
These expectation values can be obtained as functional 
derivatives of $\log \langle W(\Gamma) \rangle $ with respect to 
the path, i.e. with respect to the quark trajectories ${\bf z}_1 (t)$ 
or ${\bf z}_2 (t)$. In fact let us consider the change in 
$\langle W(\Gamma) \rangle$ induced by letting 
$ z_j^\mu (t) \rightarrow  z_j^\mu (t) + \delta  z_j^\mu (t)$ where 
$\delta  z_j^\mu  (t_{\rm i}) = \delta  z_j^\mu (t_{\rm f}) = 0$:
\begin{equation}
g \langle\!\langle F_{\mu\nu}(z_j) \rangle\!\rangle = (-1)^{j+1}
{\delta i \log \langle W(\Gamma) \rangle \over \delta S^{\mu\nu} (z_j)} \> ,
\label{e20}
\end{equation}
$$ 
\delta S^{\mu\nu} (z_j) =  
(dz_j^\mu \delta z_j^\nu - dz_j^\nu \delta z_j^\mu) \> .
$$
Varying again the path 
\begin{equation}
g^2 \left(\langle\!\langle F_{\mu\nu}(z_1) 
F_{\lambda\rho}(z_2) \rangle\!\rangle 
- \langle\!\langle F_{\mu\nu}(z_1) \rangle\!\rangle 
  \langle\!\langle F_{\lambda\rho}(z_2) \rangle\!\rangle \right)
= - i g {\delta\over \delta S^{\lambda\rho}(z_2)} 
\langle\!\langle F_{\mu\nu}(z_1) \rangle\!\rangle.
\label{e21}
\end{equation}
All contributions to the spin dependent part of the potential 
can be expressed as first and second variational derivatives of 
$\log \langle W(\Gamma) \rangle$. Therefore the whole 
quark-antiquark potential depends only on the assumed behaviour 
of $\langle W(\Gamma) \rangle$. In the next sections we will discuss 
some of these assumptions and give for each of them the explicit 
analytical expression of the potential.

\section{MINIMAL AREA LAW MODEL (MAL)}

In Ref. \cite{BCP,NC} $\langle W(\Gamma) \rangle$ 
was approximated by the sum of a perturbative part given at the 
leading order by the gluon propagator $D_{\mu\nu}$ 
and a non-perturbative part given by the value of the minimal area of 
the deformed Wilson loop of fixed contour $\Gamma$ plus a perimeter 
contribution $\cal P$:
\begin{eqnarray}
i \log \langle W (\Gamma) \rangle &=&  
i\log \langle W (\Gamma) \rangle^{\rm SR} 
+ i \log \langle W (\Gamma) \rangle^{\rm LR}
\nonumber\\
&=& - \frac{4}{3} g^2 \oint_{\Gamma}dx^{\mu}_1 
\oint_{\Gamma} dx^{\nu}_2 ~iD_{\mu \nu} (x_1-x_2)  
+ \sigma S_{\rm min} + {C\over 2} {\cal P} \>.
\label{mal}
\end{eqnarray}

Denoting by $u^{\mu}=u^{\mu}(s,t)$ the equation of 
any surface with contour $\Gamma$ ($s \in [0,1],\,
t \in [t_{\rm i},t_{\rm f}], \, 
u^0(s,t)=t, \, {\bf u}(1,t)= {\bf z}_1(t), 
\, {\bf u}(0,t)= {\bf z}_2(t) \,$) and defining 
${\bf u}_{\rm T} \equiv {\bf u} - ({\bf u}\cdot {\bf n})~{\bf n}$ 
with ${\bf n} = (\partial {\bf u}/\partial s)
|\partial {\bf u}/\partial s|^{-1}$, we can write:
\begin{eqnarray}
S_{\rm min} & = & \min \int_{t_{\rm i}}^{t_{\rm f}}
dt \, \int_0^1  ds
 \, \left[-
\left( \frac{\partial u^{\mu}}{\partial t} \frac{\partial u_{\mu}}
{\partial t} \right) \left( \frac{\partial u^{\mu}}{\partial s}
\frac{\partial u_{\mu}}{\partial s} \right) + \left(
\frac{\partial u^{\mu}}{\partial t} \frac{\partial u_{\mu}}
{\partial s} \right)^2 \right]^{\frac{1}{2}}
\nonumber\\
& = & \min \int_{t_{\rm i}}^{t_{\rm f}} dt \, \int_0^1
 ds \, \left|\frac{\partial 
{\bf u}}{\partial s } \right| \left\{ 1-\left[ \left(
\frac{\partial {\bf u}}{\partial t} \right)_{\rm T} \right]^2 \right\}^
{\frac{1}{2}} \> ,
\end{eqnarray}
which coincides with the Nambu--Goto action.
Up to the order $1/m^2$ the minimal surface  can be identified 
exactly (see App. B ref.\cite{BCP}) with the surface spanned by the 
straight-line joining $(t,{\bf z}_1(t))$ to $(t,{\bf z}_2(t))$  
with $t_{\rm i} \le t \le t_{\rm f}$. 
The generic point of this surface is 
\begin{equation}
u^0_{\min}=t \quad \quad \quad 
{\bf u}_{\min} = s~{\bf z}_1(t) + (1-s)~ {\bf z}_2(t) \>,
\label{straight}
\end{equation}
with $0\leq s \leq 1$ and ${\bf z}_1(t)$ and ${\bf z}_2(t)$ being the 
positions of the quark and the antiquark at the time $t$.
Then, the exact expression for the minimal area at the order $1 /m^2$ 
in the MAL turns out to be
\begin{eqnarray}
S_{\min} &=&
\int_{t_{\rm i}}^{t_{\rm f}} dt \,  r \int_0^1 ds \, 
[1-(s~\dot{{\bf z}}_{1 \rm T} + (1-s)~
 \dot{{\bf z}}_{2 \rm T} )^2]^{\frac{1}{2}} =
\nonumber\\
&=&  \int_{t_{\rm i}}^{t_{\rm f}} dt \,  r \,
 \left[ 1-\frac{1}{6} \left(\dot{{\bf z}}_{1 \rm T}^2 
+ \dot{{\bf z}}_{2 \rm T}^2
+ \dot{{\bf z}}_{1 \rm T} \cdot \dot{{\bf z}}_{2 \rm T} \right) 
+ \ldots \, \, \right] \>.
\end{eqnarray}
The perimeter term is given simply by
\begin{equation}
{\cal P} = \vert {\bf x}_1 - {\bf x}_2 \vert + \vert {\bf y}_1 - {\bf y}_2
\vert  + \sum_{j=1}^2 \int_{t_{\rm i}}^{t_{\rm f}} 
dt \sqrt {\dot{z}_j^\mu \dot{z}_{j\mu}} \>,
\label{per}
\end {equation}
and it is clear that we can neglect the time-independent perimeter 
contribution to the potential in the limit of big time interval 
$t_{\rm f} - t_{\rm i}$. 
By expanding also eq. (\ref{per}) at the $1/m^2$ order we have
\begin{equation}
i \log \langle W (\Gamma) \rangle^{\rm LR} 
=  \int_{t_{\rm i}}^{t_{\rm f}} dt \, \sigma r \, \left[ 1-\frac{1}{6} 
\left(\dot{{\bf z}}_{1 \rm T}^2 
+ \dot{{\bf z}}_{2 \rm T}^2
+ \dot{{\bf z}}_{1 \rm T} \cdot \dot{{\bf z}}_{2 \rm T} \right) \right]
+ {C\over 2} \sum_{j=1}^2 \int_{t_{\rm i}}^{t_{\rm f}} dt 
\left( 1-{1\over 2} \dot{z}_j^h \dot{z}_j^h \right) \>.
\end{equation}

For what concerns the perturbative part in the limit for large 
$t_{\rm f} - t_{\rm i}$ the only non-vanishing contribution 
to the Wilson loop is given by
\begin{equation}
i\log \langle W (\Gamma) \rangle^{\rm SR} 
= - \frac{4}{3} g^2 
\int_{t_{\rm i}}^{t_{\rm f}} dt_1 
\int_{t_{\rm i}}^{t_{\rm f}} dt_2 ~
{\dot z}_1^\mu(t_1) ~ {\dot z}_2^\nu(t_2) 
~iD_{\mu \nu} (z_1-z_2)  \>.
\label{sr}
\end{equation}
In the infinite time limit this expression is still gauge invariant. 
Expanding $z_2(t_2)$ around $t_1$ it is possible to evaluate explicitly 
from eq. (\ref{sr}) the short-range potential up to a given 
order in the inverse of the mass. Self-energy terms are neglected.

So, in this framework the following (MAL) static and velocity dependent
potential were obtained:
\begin{equation}
V_0 = -{4\over 3} {\alpha_{\rm s} \over r} + \sigma r + C \>,
\label{v0mal}
\end{equation}
and the explicit expressions for the potentials are:
\begin{eqnarray}
V_{\rm b}(r)&=&{8\over 9} {\alpha_{\rm s} \over r} - {1\over 9}\sigma r \>,
\quad \quad \,~
V_{\rm c}(r) = -{2\over 3} {\alpha_{\rm s} \over r} - {1\over 6}\sigma r \>,
\nonumber \\
V_{\rm d}(r)&=& -{1\over 9} \sigma r -{1\over 4} C \>, \quad \quad 
V_{\rm e}(r) = -{1 \over 6}\sigma r \>.
\label{vdmal}
\end{eqnarray}
These potentials fulfil the exact relations (\ref{relvel1}) and  
(\ref{relvel2})

Moreover by evaluating the functional derivatives for the Wilson 
loop, as given by eqs. (\ref{e20})-(\ref{e21}), we obtain also the 
spin-dependent potentials
\begin{equation}
\Delta V_{\rm a}(r) =  0 \>, \quad 
{d\over dr} V_1(r) = -\sigma  \>, \quad 
{d\over dr} V_2(r) = {4\over 3} {\alpha_{\rm s} \over r^2} \>, \quad 
V_3(r) = 4 {\alpha_{\rm s}\over r^3} \>, \quad 
V_4(r) = {32\over 3}\pi \alpha_{\rm s} \delta^3({\bf r}) \>.
\label{vsmal}
\end{equation}
These potentials reproduce the Eichten--Feinberg--Gromes results 
\cite{eichten} and fulfil the Gromes relation (\ref{grom}).  
Notice that, as a consequence of the vanishing in this model of the 
long-range behaviour of the spin-spin potential $V_{\rm SS}$ and the 
spin-orbit magnetic potential $V_{\rm LS}^{\rm MAG}$, there is no 
long-range contribution to $V_2$, $V_3$ and $V_4$.  
Instead $V_1$ has only a non-perturbative long-range contribution, 
which comes from the Thomas precession potential (\ref{vthomas}). 

The MAL model strictly corresponds to the Buchm\"uller picture 
\cite{Buch} where the magnetic field in the comoving  system is 
taken to be equal to zero. Let us first notice that the perimeter 
contributions at the $1 /m^2$ order can be simply absorbed in a 
redefinition of the quark masses $m_j \to m_j + C/2$ 
(for details see \cite{NC}). Then let us consider the moving quark 
and antiquark connected by a chromoelectric flux tube and let us 
describe the flux tube as a string with pure transverse velocity 
${\bf v}_{\rm t}$. At the classical relativistic level the system 
is described by the  flux tube Lagrangian \cite{ols2,fluxnoi}
\begin{equation}
{\cal L} = - \sum_{j=1}^2 m_j \sqrt{1- {\bf v}_j^2} 
- \sigma \int_0^r dr^\prime \sqrt{1- {\bf v}_{\rm t}^{\prime 2}} \>,
\end{equation}
with $ {\bf v}_{\rm t}^\prime  = {\bf v}_{1{\rm t}}~ {r^\prime / r}  
+ {\bf v}_{2{\rm t}} (1- {r^\prime / r} ) $.
The semirelativistic limit of this Lagrangian gives back the 
non-perturbative part of the $V_0$ and $V_{\rm VD}$ potential 
in the MAL model (notice that the minimal area law in the 
straight-line approximation is the configuration given by a straight 
flux tube) 
\footnote
{For a discussion of the relation between the two models 
in the path integral formulation see \cite{BCP}.}.
The remarkable characteristics of the obtained $V_{\rm VD}$ potential 
is the fact that it is proportional to the square of the angular 
momentum and so takes into account the energy and angular 
momentum of the string:
\begin{equation}
V_{\rm VD}^{\rm LR} = - {1\over 12 m_1 m_2}  {\sigma \over r}
({\bf L}_1 \cdot {\bf L}_2 + {\bf L}_2 \cdot {\bf L}_1 )
- \sum_{j=1}^2 { 1\over 6 m_j^2 } {\sigma \over r} {\bf L}_j^2 \>.
\label{vdflux}
\end{equation}
Finally, the non-perturbative spin-dependent part of the potential 
in this intuitive flux tube picture simply comes from the Buchm\"uller 
ansatz that the  chromomagnetic field is zero in the comoving framework 
of the flux tube.

We notice that even if $V_1$ seems to arise from an effective Bethe--Salpeter 
kernel which is a scalar and depends only on the momentum transfer, 
a simple convolution kernel cannot reproduce the correct velocity dependent 
potential (\ref{vdflux}) or equivalently (\ref{vdmal}) \cite{report}.
Nevertheless the behaviour (\ref{vdflux}) seems to be important to reproduce 
the spectrum \cite{Gupta,Durand,Lagae,Dubin,letnoi,Olsson}.

\section{STOCHASTIC VACUUM MODEL (SVM)}

The SVM (see \cite{Sinp,DoSi} and for a review \cite{DoSi2}) in the context 
of heavy quark bound state gives a justification of the MAL model 
avoiding the artificial splitting of the Wilson loop in a 
perturbative and a non-perturbative part. It reproduces the flux tube 
distribution measured on the lattice \cite{rueter}. 
Moreover it allows to go beyond the MAL model in a systematic way 
(e.g. with the so-called perturbation theory in non-perturbative 
background \cite{back}). The whole non-perturbative physics is 
factorized in some correlation function which can be 
calculated on the lattice.

The starting point is to express the Wilson loop average 
$\langle W(\Gamma) \rangle$ via the non-Abelian Stokes 
theorem \cite{Stokes,Si} in terms of an integral over a surface 
$S$ enclosed by the contour $\Gamma$, 
and then to perform a cluster expansion \cite{Kampen}.
In order to allow lattice calculations all these quantities 
are given in the Euclidean metric. Some care must be payed 
in converting it in the Minkowskian metric before putting 
in eq. (\ref{potential}).
\begin{eqnarray}
\langle W(\Gamma) \rangle &=& 
\left\langle {\rm P} \>
\exp \left( ig \int_S dS_{\mu\nu}(u) F_{\mu\nu}(u,x_0) \right) \right\rangle 
\label{stokes}\\ 
&=& 
 \exp \sum_{j=1}^{\infty} {(ig)^j\over j!} 
\int_S dS_{\mu_1\nu_1}(u_1) \dots 
\nonumber\\
&~& \quad\quad\quad 
\int_S dS_{\mu_j\nu_j} (u_j) 
\langle F_{\mu_1\nu_1}(u_1,x_0) \dots F_{\mu_j\nu_j}(u_j,x_0)
\rangle_{\rm cum}  \>.
\label{cluster}
\end{eqnarray}
The cumulants $\langle ~\> \rangle_{\rm cum}$ are defined 
in terms of average values over the gauge fields $\langle ~\> \rangle$: 
\begin{equation}
\langle F(1)\rangle_{\rm cum} = \langle F(1) \rangle \>, \quad\quad 
\langle F(1) F(2) \rangle_{\rm cum} = 
\langle F(1) F(2)\rangle  - \langle F(1) \rangle \langle F(2)\rangle
\>, ~~ \dots 
\nonumber
\end{equation}
and  ${\rm P \>}F_{\mu,\nu}(u,x_0) \equiv 
{\rm P \>} \exp \left[i g \int_{x_0}^u dx^\mu A_\mu (x) \right] 
F_{\mu\nu}(u) \exp \left[i g \int^{x_0}_u dx^\mu A_\mu (x) \right]$
where $x_0$ is an arbitrary reference point on the surface $S$ 
appearing in the non-Abelian Stokes theorem (\ref{stokes}).
In general each cumulant depends on $S$ and on 
$x_0$, but, as the left-hand side of eq. (\ref{stokes}) does not, 
it is expected that in the full resummation of all the 
cumulants (right-hand side of eq. (\ref{cluster})) this dependence 
will disappear \cite{Si}. To minimize the required cancellations 
$S$ is chosen to be the minimal area surface.

Equation (\ref{cluster}) is exact. The first cumulant 
vanishes trivially. The second cumulant gives the first non-zero 
contribution to the cluster expansion  (\ref{cluster}). In the SVM 
one assumes that in the context of heavy quark bound states 
higher cumulants can be neglected and the 
second cumulant dominates the cluster expansion, or, in other words, 
that the vacuum fluctuations are of a Gaussian type:
\begin{equation}
\log \langle W(\Gamma) \rangle  = -{g^2 \over 2} 
\int_S dS_{\mu\nu}(u) \int_S dS_{\lambda\rho} (v)  
\langle F_{\mu\nu}(u,x_0)  F_{\lambda\rho}(v,x_0) \rangle_{\rm cum}
\> .
\label{svm}
\end{equation}
Neglecting the  dependence on $x_0$ and on the arbitrary curves 
connecting $x_0$ with $u$ and $v$ which seems to be relegated to higher 
correlators, the Lorentz structure of the bilocal cumulant 
implies that it can be expressed as \cite{Sinp}:
\begin{eqnarray}
\langle F_{\mu\nu}(u,x_0)  F_{\lambda\rho}(v,x_0) \rangle_{\rm cum}
&=& \langle F_{\mu\nu}(u,x_0) F_{\lambda\rho}(v,x_0) \rangle \> 
\nonumber\\
&=& {\beta\over g^2} \Bigg\{
(\delta_{\mu\lambda}\delta_{\nu\rho} - 
\delta_{\mu\rho}\delta_{\nu\lambda})D\left( (u-v)^2 \right) 
\nonumber\\
&+& {1\over 2}\left[
{\partial\over\partial u_\mu}\left( (u-v)_\lambda\delta_{\nu\rho} 
- (u-v)_\rho\delta_{\nu\lambda} \right) \right.
\nonumber\\
&~&\quad\quad\quad \left.
+ {\partial\over\partial u_\nu}\left( (u-v)_\rho\delta_{\mu\lambda} 
- (u-v)_\lambda\delta_{\mu\rho} \right) \right] D_1 \left( (u-v)^2 \right)
\Bigg\}
\label{cum2}\\
\beta&\equiv&{g^2\over 36} 
{\langle {\rm Tr \>} F_{\mu\nu}(0) F_{\mu\nu}(0) 
\rangle \over D(0) + D_1(0)} \> .
\nonumber
\end{eqnarray}
Eq. (\ref{svm}) and (\ref{cum2}) define the SVM for heavy quarks. 
The correlator functions $D$ and $D_1$ are unknown. 
The perturbative part of $D_1$, which is expected to be dominant 
in the short-range behaviour, can be obtained by means of the 
standard perturbation theory:
\begin{equation}
D_1^{\rm pert} (x^2) = {16 \alpha_{\rm s} \over 3\pi}{1\over x^4} 
+ ~{\rm higher ~~orders} \>. 
\label{D1pert}
\end{equation}
Instead the only information which we know about the non-perturbative 
contributions to $D$ and $D_1$ come from lattice simulations.
A good parametrization of the long-range behaviour 
of the bilocal correlators seems to be \cite{lattice2,BaSi}:
\begin{eqnarray}
\beta~D^{\rm LR}(x^2) &=& d~e^{-\delta|x|} \>,
\quad \> \delta = (1 \pm 0.1) ~{\rm GeV} \>, 
\quad d = 0.073 ~{\rm GeV}^4 \>,
\label{Dlr}\\
\beta~D_1^{\rm LR}(x^2) &=& d_1~e^{-\delta_1|x|} \>,
\quad \delta_1 = (1 \pm 0.1) ~{\rm GeV} \>, 
\quad d_1 = 0.0254 ~{\rm GeV}^4 \>,
\label{D1lr}
\end{eqnarray}

Up to order $1/m^2$ the minimal area 
surface can be identified, as in the previous section, with   
the straight-line surface (\ref{straight}). 
In particular, since $dS_{\mu\nu}(u)$ $\equiv$ 
$dt~ds~ $  $\partial u_\mu(t,s)/ \partial t ~~$ 
$\partial u_\nu(t,s)/ \partial s$, we have:
\begin{eqnarray}
dS_{4j}(u)     &=& dt~ds~ r_j(t)       \>;
\nonumber\\
dS_{ij}(u)     &=& dt~ds~ \left( s~ {\dot z}_{i1}(t) 
+ (1-s)~ {\dot z}_{i2}(t) \right)~r_j(t) \>.
\nonumber
\end{eqnarray}

From (\ref{svm}) and (\ref{cum2}) and taking in account 
(\ref{straight}) we have calculated explicitly 
$\log \langle W(\Gamma) \rangle$. Considering time interval 
much larger than the typical correlation length of $D$ and $D_1$, 
up to order $1/m^2$ we have (for details see the appendix):
\begin{eqnarray}
V_0(r) &=&  \beta \int_{-\infty}^{+\infty} d\tau \left\{ 
\int_0^r d\lambda (r-\lambda)~D(\tau^2+\lambda^2) 
+ \int_0^r d\lambda {\lambda\over 2}D_1(\tau^2+\lambda^2) \right\} \>,
\label{v0svm}\\
V_{\rm b}(r) &=&  {\beta\over 6} \int_{-\infty}^{+\infty} d\tau \left\{ 
\int_0^r d\lambda 
\left( -{2\over 3}r - {\lambda^2\over r} + {8\over 3}{\lambda^3\over r^2} 
-3{\tau^2\over r} \right) ~D(\tau^2+\lambda^2) \right.
\nonumber\\
&~&\quad\quad \left.
+ \int_0^r d\lambda 
\left( -{3\over 2} {\lambda^2 \over r} 
+ {3\over 2}{\tau^2\over r} \right) D_1(\tau^2+\lambda^2) 
+ {r^2\over 2} D_1(\tau^2+r^2)
\right\} \>,
\label{vbsvm}\\
V_{\rm c}(r) &=&  {\beta\over 2} \int_{-\infty}^{+\infty} d\tau \left\{ 
\int_0^r d\lambda 
\left( -{r\over 3} - 2 {\lambda^2\over r} + {4\over 3}{\lambda^3\over r^2} 
\right) ~D(\tau^2+\lambda^2) \right.
\nonumber\\
&~&\quad\quad \left.
- {r^2\over 2} D_1(\tau^2+r^2)
\right\} \>,
\label{vcsvm}\\
V_{\rm d}(r) &=&  {\beta\over 6} \int_{-\infty}^{+\infty} d\tau \left\{ 
\int_0^r d\lambda 
\left( -{2\over 3}r + {3\over 2}\lambda 
+{1\over 2} {\lambda^2\over r} - {4\over 3}{\lambda^3\over r^2} 
+{3\over 2}{\tau^2\over r} \right) ~D(\tau^2+\lambda^2) \right.
\nonumber\\
&~&\quad\quad \left.
+ \int_0^r d\lambda 
\left( -{3\over 4}\lambda +{3\over 4} {\lambda^2 \over r} 
- {3\over 4}{\tau^2\over r} \right) D_1(\tau^2+\lambda^2) 
\right\} \>,
\label{vdsvm}\\
V_{\rm e}(r) &=&  {\beta\over 2} \int_{-\infty}^{+\infty} d\tau 
\int_0^r d\lambda 
\left( -{r\over 3} +{\lambda^2\over r} - {2\over 3}{\lambda^3\over r^2} 
\right) ~D(\tau^2+\lambda^2) \>.
\label{vesvm}
\end{eqnarray}
Result (\ref{v0svm}) was found in \cite{Sinp}, whereas 
(\ref{vbsvm})-(\ref{vesvm}) are new. We note that these expressions for 
the potentials $V_0$ and $V_{\rm b}$, ..., $V_{\rm e}$ satisfy 
identically the Barchielli--Brambilla--Prosperi relations (\ref{relvel1}) 
and (\ref{relvel2}). Of particular interest seems to be the potential 
$V_{\rm e}$ that has only non-perturbative contributions in the bilocal 
approximation.

To evaluate the spin dependent part of the potential, the only 
terms which we need are those with one and two field strength 
insertions (taking in account that
$\langle\!\langle D^{\nu} F_{\nu\mu}(x) \rangle\!\rangle  = 
\partial_\nu \langle\!\langle F_{\nu\mu}(x) \rangle\!\rangle$). 
By means of eq. (\ref{e20}) and (\ref{e21}) and (\ref{svm}):
\begin{eqnarray}
g \langle\!\langle F_{0l}(z_j) \rangle\!\rangle &=& 
\beta r_l \int_{-\infty}^{+\infty} d\tau \left\{ 
\int_0^r d\lambda {1\over r} D(\tau^2+\lambda^2)
+{1\over 2} D_1(\tau^2+r^2) \right\} \>,
\nonumber\\
g \langle\!\langle F_{il}(z_1) \rangle\!\rangle &=&
\beta \left( {\dot z}_{l1}r_i - {\dot z}_{i1}r_l \right) 
\int_{-\infty}^{+\infty} d\tau 
\int_0^r d\lambda {1\over r} \left(1- {\lambda\over r} \right)
D(\tau^2+\lambda^2)  
\nonumber\\
&+& 
\beta \left( {\dot z}_{l2}r_i - {\dot z}_{i2}r_l \right) 
\int_{-\infty}^{+\infty} d\tau \left\{ 
\int_0^r d\lambda {\lambda \over r^2} D(\tau^2+\lambda^2) 
+{1\over 2} D_1(\tau^2+r^2) \right\} \>,
\nonumber\\
g \langle\!\langle F_{il}(z_2) \rangle\!\rangle &=&
\beta \left( {\dot z}_{l2}r_i - {\dot z}_{i2}r_l \right) 
\int_{-\infty}^{+\infty} d\tau 
\int_0^r d\lambda {1\over r} \left(1- {\lambda\over r} \right)
D(\tau^2+\lambda^2)  
\nonumber\\
&+& 
\beta \left( {\dot z}_{l1}r_i - {\dot z}_{i1}r_l \right) 
\int_{-\infty}^{+\infty} d\tau \left\{ 
\int_0^r d\lambda {\lambda \over r^2} D(\tau^2+\lambda^2) 
+{1\over 2} D_1(\tau^2+r^2) \right\} \>,
\nonumber
\end{eqnarray}
\begin{eqnarray}
&~&
g^2 \left(\langle\!\langle F_{\mu\nu}(z_1) 
F_{\lambda\rho}(z_2) \rangle\!\rangle 
- \langle\!\langle F_{\mu\nu}(z_1) \rangle\!\rangle 
  \langle\!\langle F_{\lambda\rho}(z_2) \rangle\!\rangle \right) = 
\quad\quad\quad\quad\quad\quad\quad\quad\quad
\nonumber\\
&~&\quad\quad\quad\quad\quad
\beta (\delta_{\mu\lambda}\delta_{\nu\rho} - 
\delta_{\mu\rho}\delta_{\nu\lambda})
\left( D(\tau^2+r^2)+D_1(\tau^2+r^2) \right) 
\nonumber\\
&~&\quad\quad\quad\quad
+ \beta\left(
  r_\mu r_\lambda \delta_{\nu\rho} - r_\mu r_\rho \delta_{\nu\lambda} 
+ r_\nu r_\rho \delta_{\mu\lambda} - r_\nu r_\lambda \delta_{\mu\rho} 
\right) {\partial \over \partial \tau^2} D_1(\tau^2 + r^2) \>, 
\quad \quad r_4 \equiv \tau = t_1-t_2\>.
\nonumber 
\end{eqnarray}
In this way we obtain the following expressions
for the spin-dependent potentials in the SVM (confirming the results 
obtained in \cite{Sinp} with a different derivation):
\begin{eqnarray}
\Delta V_{\rm a}(r) &=& {\hbox{self-energy terms}} \>, 
\label{vasvm} \\
{d\over dr}V_1(r) &=&  
- \beta \int_{-\infty}^{+\infty} d\tau 
\int_0^r d\lambda \left(1-{\lambda \over r}\right)~D(\tau^2+\lambda^2) \>,
\label{v1svm}\\
{d \over dr}V_2(r) &=&  
\beta \int_{-\infty}^{+\infty} d\tau \left\{
\int_0^r d\lambda {\lambda \over r}~D(\tau^2+\lambda^2) 
+ {1\over 2} r D_1(\tau^2+r^2) \right\} \>,
\label{v2svm}\\
V_3 (r) &=&  
- \beta \int_{-\infty}^{+\infty} d\tau ~ 
r^2 {\partial \over \partial \tau^2} D_1(\tau^2+r^2) \>,
\label{v3svm}\\
V_4 (r) &=&  
\beta \int_{-\infty}^{+\infty} d\tau ~\left\{
3 D(\tau^2+r^2)  + 3 D_1(\tau^2+r^2)
+ 2 r^2 {\partial \over \partial \tau^2} D_1(\tau^2+r^2) \right\} \>.
\label{v4svm}
\end{eqnarray}
Potentials (\ref{v0svm}), (\ref{v1svm}) and (\ref{v2svm}) 
satisfy identically the Gromes relation (\ref{grom}).
An application of the spin potentials to the $b\bar{b}$ and 
$c\bar{c}$ spectrum, with a discussion on the different type 
of parametrization of the correlation functions, can be found in 
\cite{BaSi,BaYu}.

In the short-range behaviour ($r\to 0$), assuming that 
all the relevant contributions come from the perturbative 
part of $D_1$ (\ref{D1pert}) eqs.(\ref{v0svm})-(\ref{vesvm}) and 
(\ref{v1svm})-(\ref{vasvm}) exactly reproduce (after subtracting 
the self-energy contributions) the 
$\alpha_{\rm s}$-depending part of eqs. (\ref{v0mal}), 
(\ref{vdmal}) and (\ref{vsmal}) of the MAL model. 
We observe that no gauge choice is necessary in this approach, 
which is manifestly gauge-invariant. 
Moreover we note that the short-range behaviour of the 
$D_1$ correlator is not {\it ad hoc} but emerges straightforwardly 
from the comparison with the $\alpha_{\rm s}$ expansion of the Wilson loop.

In the long-range behaviour ($r\to\infty$):
\begin{equation}
V_0(r) = \sigma_2 r  + {1\over 2}C_2^{(1)} - C_2 \> , \quad
{d\over dr}V_1(r) = -\sigma_2 + {C_2 \over r} \>, \quad
{d\over dr}V_2(r)  =  {C_2 \over r}  \>,
\label{vsosvm}
\end{equation} 
$V_3$ and $V_4$ fall off exponentially and 
\begin{eqnarray}
\Delta V_{\rm a}(r) &=& {\hbox{self-energy terms}} \>,
\nonumber\\
V_{\rm b}(r) &=& -{1\over 9} \sigma_2 r -{2\over 3}{D_2 \over r} 
+ {8\over 3}{E_2\over r^2}\>,
\quad\quad\quad\quad\quad\quad\quad\>~
V_{\rm c}(r)  = -{1\over 6} \sigma_2 r -{D_2\over r} + {2\over 3}
{E_2\over r^2}\>,
\label{vvsvm}\\
V_{\rm d}(r) &=& -{1\over 9} \sigma_2 r + {1\over 4}C_2 - {1\over 8}C_2^{(1)} 
+ {1\over 3}{D_2\over r} - {2\over 9}{E_2 \over r^2}\>,
\quad
V_{\rm e}(r)  = -{1\over 6} \sigma_2 r + {1\over 2}{D_2\over r} 
- {1\over 3}{E_2\over r^2}\>,
\nonumber
\end{eqnarray}
with
\begin{eqnarray}
\sigma_2 &\equiv& \beta \int_{-\infty}^{+\infty} d\tau 
\int_0^\infty d\lambda ~D(\tau^2+\lambda^2) \>, 
\nonumber\\
C_2 &\equiv& \beta \int_{-\infty}^{+\infty} d\tau 
\int_0^\infty d\lambda ~\lambda ~D(\tau^2+\lambda^2) \>, 
\quad\quad ~
C_2^{(1)} \equiv \beta \int_{-\infty}^{+\infty} d\tau 
\int_0^\infty d\lambda ~\lambda ~D_1(\tau^2+\lambda^2) \>,
\nonumber\\
D_2 &\equiv& \beta \int_{-\infty}^{+\infty} d\tau 
\int_0^\infty d\lambda ~\lambda^2 ~D(\tau^2+\lambda^2) \>, 
\nonumber\\
E_2 &\equiv& \beta \int_{-\infty}^{+\infty} d\tau 
\int_0^\infty d\lambda ~\lambda^3 ~D(\tau^2+\lambda^2) \>.
\nonumber
\end{eqnarray}
By means of parametrization (\ref{Dlr})-(\ref{D1lr}) we have 
\begin{eqnarray}
\sigma_2 &=& {\pi d\over \delta^2}\simeq 0.2 ~{\rm GeV^2} \>, \quad
C_2={4 d\over \delta^3}\simeq 0.3 ~{\rm GeV} \>, 
\nonumber\\
C_2^{(1)} &=& {4 d_1\over \delta_1^2}\simeq 0.1 ~{\rm GeV} \>, \quad
D_2={3\pi d\over \delta^4}\simeq 0.7 \>, \quad\> 
E_2= {32 d \over \delta^5} \simeq 2.3 ~{\rm GeV}^{-1} \>, 
\nonumber 
\end{eqnarray}
and 
\begin{eqnarray}
V_3(r) &=& 1.2(6) ~ d_1 \sqrt{\delta_1}  ~ r^{3\over 2} e^{-\delta_1 r} \>,
\nonumber \\
V_4(r) &=& 1.2(6) ~{6 d\over \sqrt{\delta}} ~ r^{1\over 2} e^{-\delta r} 
+ 1.2(6) ~{6 d_1\over \sqrt{\delta_1}} ~ r^{1\over 2} e^{-\delta_1 r} 
- 2.5(2) ~d_1 \sqrt{\delta_1} ~ r^{3\over 2} e^{-\delta_1 r} \>.
\nonumber 
\end{eqnarray}

Identifying $\sigma_2$ with $\sigma$ and $C_2^{(1)}/2 - C_2$ with $C$ 
then at the leading order in $r\to\infty$ the spin-dependent and 
velocity-dependent SVM potentials reproduce the long-range behaviour 
of the potentials (\ref{v0mal}), (\ref{vdmal}) and (\ref{vsmal}) 
in the MAL model. Notice that the constant terms in the static and velocity 
dependent potentials turn out in the same combination as necessary 
to be reabsorbed in a redefinition of the quark masses. Some differences 
emerge at the next orders. In the SVM the magnetic contribution to the 
spin-orbit potential (which we called $V_{\rm LS}^{\rm MAG}$ 
in Sec. 2) is not exactly zero in the long-range behaviour but gives some 
$1/r$ corrections. For this reason the potential $dV_2/dr$ does not 
vanish and the potential $dV_1/dr$ presents a $1/r$ correction 
to the Thomas precession term. Notice, also, that in the SVM 
the tensor potential $V_3$ and the spin-spin potential $V_4$ 
are exponentially decreasing with the distance $r$ but not 
identically zero as in the MAL model. In the next section we will see 
how the Dual QCD model is able to reproduce this behaviour. Finally a very 
rich structure of entirely non-perturbative $1/r$ and $1/r^2$ corrections 
emerges in the velocity dependent part of the potential.
A lattice study of this kind of contributions is in progress 
\cite{Bali} and in the light of eqs. (\ref{vvsvm}) 
should give an interesting check on the validity of the 
stochastic vacuum approach in the velocity dependent 
sector of the potential and possibly some new indications 
on the behaviour of the correlator function $D$. 
A last comment on the fact that $\Delta V_{\rm a}$ is not $r$ dependent.
This is a direct consequence of the bilocal approximation 
which we have adopted. In principle nothing prevents us from the existence 
of $r$ dependent contributions coming from higher order cumulants. 
We think it will be an important task to estimate such kind of contributions 
and compare it with lattice results (for a more detailed discussion 
see \cite{BVa}).

\section{DUAL QCD (DQCD)}

The duality assumption that the long distance physics of a Yang--Mills
theory depending upon strong coupled gauge potentials $A_\mu$ is
the same as the long distance physics of the dual theory describing 
the interactions of weakly coupled dual potentials 
${\cal C}_\mu\equiv \sum_{a=1}^8 C_\mu^a \lambda_a / 2$ and monopole 
fields ${\cal B}_i\equiv \sum_{a=1}^8 B_i^a \lambda_a/2$, 
forms the basis of DQCD \cite{dual}\footnote{
The name Dual QCD has historical reasons, but can give rise 
to some confusions. We emphasize that the duality assumption concern 
only the long distance physics of a strongly coupled Yang--Mills theory  
as the gluonic sector of QCD. }. 
The model is constructed as a concrete realization of 
the Mandelstam--t'Hooft \cite{Mandelstam} dual superconductor mechanism 
of confinement. Indeed, the explicit form of the Lagrangian expressed 
in terms of the dual potentials is not known in a non-Abelian Yang--Mills 
theory. Since the main interest is solving such a theory in the 
long-distance regime, the Lagrangian ${\cal L}_{\rm eff}$ is explicitly 
constructed as the minimal dual gauge invariant extension of a quadratic 
Lagrangian with the further requisite to give a mass to the dual gluons 
(and to the monopole fields) via a spontaneous symmetry breaking of the 
dual gauge group.

We denote by $\langle W_{\rm eff} (\Gamma)\rangle$ the average 
over the fields of the Wilson loop of the dual theory \cite{dualin}:
\begin{equation}
\langle W_{\rm eff} (\Gamma) \rangle= 
{\int {\cal D} C_\mu {\cal D} B {\cal D} B_3
e^{i \int dx [ {\cal L}_{\rm eff} (G_{\mu\nu}^{\rm S}) 
+ {\cal L}_{\rm GF} ] }
\over \int {\cal D} C_\mu {\cal D} B {\cal D} B_3
e^{i \int dx [ {\cal L}_{\rm eff} (G_{\mu\nu}^{\rm S}=0) 
+ {\cal L}_{\rm GF} ] } } \>,
\label{weff}
\end{equation}
where ${\cal L}_{\rm GF}$ is a gauge fixing term and 
the effective dual Lagrangian in presence of quarks
is given by 
\begin{equation} 
{\cal L}_{\rm eff}(G_{\mu\nu}^{\rm S}) = 
2~{\rm Tr} 
\left\{ - {1\over 4} {\cal G}^{\mu\nu} {\cal G}_{\mu\nu}
+ {1\over 2} ({\cal D}_\mu {\cal B}_i)^2  \right\} 
- U({\cal B}_i) \>.
\label{leff}
\end{equation}
$U({\cal B}_i)$ is the Higgs potential with a minimum at a 
non-zero value  ${\cal B}_{01} = B_0 \lambda_7$, 
${\cal B}_{02} = - B_0 \lambda_5$ and ${\cal B}_{03} = B_0 \lambda_2$. 
It was also taken $B_1=B_2=B$.
In (\ref{weff}) we have taken the dual potential 
 proportional to the hypercharge matrix ${\cal C}_\mu =C_\mu Y$ 
\footnote{Doing so ${\cal L}_{\rm eff}$ without quark sources 
generates classical equations of motion with solutions dual to the 
Abrikosov--Nielsen--Olesen magnetic vortex solutions in a 
superconductor \cite{dual,dualin}.}.
Moreover 
\begin{eqnarray}
{\cal D}_\mu {\cal B}_i &=& \partial_\mu {\cal B}_i 
+ i e [ {\cal C}_\mu, {\cal B}_i] \>, \quad\quad  e\equiv{2\pi \over g} \>,
\\
{\cal G}_{\mu\nu} &=& \left(
\partial_\mu C_\nu - \partial_\nu C_\mu 
+ G_{\mu\nu}^{\rm S} \right) Y \>, 
\\
G_{\mu\nu}^{\rm S}(x) &\equiv& g
\varepsilon_{\mu\nu\alpha\beta} \int ds \int d\tau 
{\partial y^\alpha \over \partial s} {\partial y^\beta\over \partial \tau}
\delta(x-y(s,\tau)) \>,
\label{dft}
\end{eqnarray}
and $y(s,\tau)$ is a world sheet with boundary $\Gamma$ swept out by the
Dirac string.  Notice that dual potentials couple to electric color 
charge like  ordinary potentials couple to monopoles \cite{dual,Dirac}.

The functional integral $\langle W_{\rm eff} (\Gamma) \rangle$ 
determines in DQCD the same physical quantity as 
$\langle W(\Gamma) \rangle$ in QCD. The coupling 
in ${\cal L}_{\rm eff} (G_{\mu\nu}^{\rm S})$ of the dual potentials 
to the Dirac string plays the role in the expression (\ref{weff}) of 
the Wilson loop $W(\Gamma)$ of QCD (\ref{wgamma}) in 
$\langle W(\Gamma) \rangle$. The assumption that the dual theory 
describes the long distance $Q \bar Q$ interaction in QCD 
then takes the form:
\begin{equation}
\langle W(\Gamma) \rangle  = \langle W_{\rm eff} (\Gamma) \rangle  \>, 
\quad {\rm for ~large ~loops ~\Gamma}.
\label{ass}
\end{equation}
Large loop means that the size $R$ of the loop is
large compared to the inverse mass ($M^{-1} \simeq$ (600 MeV)$^{-1}$)
of the Higgs particle (monopole field). Furthermore, since the dual 
theory is weakly coupled at large distances, we can evaluate 
$\langle W_{\rm eff} (\Gamma) \rangle$ via a semiclassical expansion to 
which the classical configuration of dual potentials and monopoles gives 
the leading contribution.  This then allows us to picture heavy quarks 
(or constituent quarks) as sources of a long distance classical field of
dual gluons determining the heavy quark potential. We mention here 
that DQCD reproduces the lattice flux tube distribution \cite{baker}. 

Eq. (\ref{ass}) defines the DQCD model for heavy quark 
bound states. Replacing $\langle W(\Gamma) \rangle$ by 
$\langle W_{\rm eff}(\Gamma) \rangle$ in eq. (2.1) we obtain expressions 
for $V_0$ and $V_{\rm VD}$ and by considering the variation 
in $\langle W_{\rm eff} (\Gamma) \rangle$ produced by the change
$G_{\mu\nu}^{\rm S} (x) \rightarrow G_{\mu\nu}^{\rm S} (x) 
+ \delta G_{\mu\nu}^{\rm S} (x)$ we obtain also the field averages  
in terms of dual quantities:
\begin{equation}
g \langle\!\langle F_{\mu \nu} (z_j)\rangle\!\rangle=
(-1)^{j+1} {\delta i \log \langle W_{\rm eff} (\Gamma) \rangle 
\over \delta S^{\mu\nu} (z_j)} = {4\over 3} g \langle\!\langle \hat
G_{\mu\nu} (z_j) \rangle\!\rangle_{\rm eff}= (-1)^{j+1} {g\over 2}\,
\varepsilon_{\mu \nu \lambda \sigma} 
{\delta i \log \langle W_{\rm eff}(\Gamma) \rangle
\over \delta G_{\lambda \sigma}^{\rm S}(z_j) }.
\end{equation}
This gives a correspondence between local quantities in
the Yang--Mills theory and in the dual theory. A similar expression can be 
obtained for the double field strength insertion in (\ref{potential}).

The weak coupling of the dual theory permits the explicit evaluation 
of $\langle W_{\rm eff}(\Gamma) \rangle$ by means of the classical 
approximation. Hence we have
\begin{equation}
i \log \langle W_{\rm eff}(\Gamma) \rangle 
= - \int dx ~{\cal L}_{\rm eff} (G_{\mu\nu}^{\rm S}) \>,
\end{equation}
with ${\cal L}_{\rm eff} (G_{\mu\nu}^S)$  evaluated
at the solution of the classical equations of motion:
\begin{eqnarray}
&~& \partial^\alpha (\partial_\alpha C_\beta - \partial_\beta C_\alpha) = 
- \partial^\alpha G_{\alpha \beta}^{\rm S} + j_\beta^{\rm MON} \>,
\label{nl1}\\
&~& (\partial_\mu + i e C_\mu)^2 B = - {1\over 4} {\delta U\over\delta B} \>,
\label{nl2}\\
&~& \partial^2 B_3 = - {1\over 4} {\delta U\over\delta B_3} \>,
\label{nl3}
\end{eqnarray}
where $j_\mu^{\rm MON}= - 6~e^2 C_\mu B^2$ is the monopole current.
The Dirac string is chosen to be a straight line connecting 
$Q$ and $\bar{Q}$ since this is the configuration having the 
minimum field energy. As a consequence of the classical approximation 
all quantities in brackets are replaced by their classical values 
$\langle\!\langle G_{\mu\nu}(x) \rangle\!\rangle_{\rm eff} 
= G_{\mu\nu} (x)$ which are obtained by solving numerically 
the non-linear equations (\ref{nl1})-(\ref{nl3}). 
An interpolation of the numerical results for the potentials 
can be found in \cite{dual} (in particular in the first of these 
references it is possible to find also an application of the 
DQCD potentials to the heavy quarkonia spectrum). 
In the following we will give and discuss 
only the large distances limit of these potentials.

In the long-range behaviour ($r \to \infty$) the interpolation 
of Ref. \cite{dual} gives
\begin{eqnarray}                                           
V_0 (r)&=& \sigma r - 0.646 \sqrt{\sigma \alpha_{\rm s}}   \>,
\label{v0dqcd} \\
{d \over dr} V_1(r) &=& -\sigma 
+ { 0.681 \over r} \sqrt{\sigma \alpha_{\rm s}} \>, 
\label{v1dqcd}\\
{d \over dr} V_2(r) &=& {0.681\over r} \sqrt{\sigma \alpha_{\rm s}}\>,
\label{v2dqcd} 
\end{eqnarray}
and
\begin{eqnarray}
V_{\rm b}(r) &=& - 0.097 ~\sigma r -0.226 \sqrt{\sigma \alpha_{\rm s}} \>,
\label{vbdqcd}\\ 
V_{\rm c}(r)  &=& - 0.146 ~\sigma r -0.516 \sqrt{\sigma \alpha_{\rm s}} \>,
\label{vcdqcd}\\ 
V_{\rm d}(r) &=& -0.118 ~\sigma r+ 0.275\sqrt{\sigma \alpha_{\rm s}} \>,
\label{vddqcd}\\ 
V_{\rm e}(r)  &=& - 0.177 ~\sigma r + 0.258\sqrt{\sigma \alpha_{\rm s}}  \>.
\label{vedqcd} 
\end{eqnarray}
For the spin-spin interaction and for large distances it is possible to 
give the exact analytical expression of the potentials:
\begin{eqnarray}
V_{3}(r) &=& {4\over 3} \alpha_{\rm s} 
~\left( M^2 + {3\over r} M + {3\over r^2} \right) ~{e^{-Mr} \over r} \>,
\label{v3dqcd}\\ 
V_{4}(r) &=& {4\over 3} \alpha_{\rm s} ~M^2 {e^{-Mr} \over r} \>.
\label{v4dqcd}
\end{eqnarray}
While $V_{\rm a}$ is, at the moment, lacking either in an analytical 
or a numerical evaluation, and is formally given by \cite{dualin}:
\begin{equation}
\Delta V_{\rm a} (r) = 
- \Delta V_0^{\rm NP} (r) 
- {4\over 3} g^2 \sum_{j=1}^3{d^2\over dx_jdx^\prime_j}
G^{\rm NP} ({\bf x}, {\bf x}^\prime) 
\Bigg \vert_{{\bf x} = {\bf x}^\prime = {\bf z}_i},
\label{vadqcd}
\end{equation}
where the first term is the color electric contribution 
to $V_{\rm a}$ ($V_0^{\rm NP} (r)$ is the non-perturbative part of 
the static potential, so that $V_{\rm a}$ is determined by
the non-perturbative gluodynamics) and the second is the color
magnetic contribution. $G^{\rm NP}$ satisfies the equation
\begin{equation}
(-\Delta + 6~ e^2 B^2) G^{\rm NP} 
= - {6~ e^2 B^2 ({\bf x})\over 4\pi|{\bf x} - {\bf x}'|} \>.
\label{nl4}
\end{equation}

The potentials depend on the two free parameters $\alpha_{\rm s}= \pi/ e^2$ 
and $\sigma$. In \cite{dual} the values
\begin{equation}
\sigma=0.18 ~{\rm GeV}^2 \>,
\quad \quad \quad \quad\quad \alpha_{\rm s} = 0.39 \>,
\nonumber
\end{equation}
were used. The dual gluon mass $M$ is related to this two 
parameters and is approximately given by:
\begin{equation}
M^2 \simeq {\pi\over 4} {\sigma \over \alpha_{\rm s}} \simeq 
(600 \> {\rm MeV})^2 \>.
\label{monopole}
\end{equation}

Finally we observe that all these potentials satisfy 
identically the Gromes relation (\ref{grom}) and the equivalent 
relations for the velocity dependent potentials 
(\ref{relvel1}) and (\ref{relvel2}). 

From the comparison of eqs. (\ref{v0dqcd})-(\ref{v1dqcd}) with 
(\ref{vsosvm}) if follows immediately that in the long-range behaviour 
the static and the spin-orbit potentials coincide completely in 
DQCD and in the SVM.  Very important seems to be the agreement, 
which we note here for the first time, between the $1/r$ corrections 
in the two models. These corrections come from the physics beyond 
the minimal area law assumption and in fact there are not present in 
the MAL model (see (\ref{vsmal})). The coefficient of the $1/r$ contribution 
in $d V_1/dr $ and $d V_2/dr$ is the same in DQCD and SVM and in both cases 
compatible with the constant term in the static potential $V_0$. 
The little difference between the constant in $d V_1/dr $, $d V_2/dr $ 
and $V_0$ can be understood in the SVM language as due to the presence 
of the small positive constant $C_2^{(1)}/2$. 
The spin-spin interaction falls off exponentially in both the models.
In DQCD the behaviour is like a Yukawa interaction, while 
eqs. (\ref{v3svm})-(\ref{v4svm}) seem not to reproduce this 
behaviour at least with parametrization (\ref{Dlr})-(\ref{D1lr}).
This is, at the moment, an important disagreement because one 
of the basic feature of DQCD is that the magnetic interaction 
(like in the spin-spin case) is carried by a massive particle.
Differences arise for large distances also in the velocity dependent 
sector and with respect to the MAL model. The factors in front of 
the $\sigma r$ leading contributions to $V_{\rm b}$, ..., $V_{\rm e}$ 
are slightly different from those of eqs. (\ref{vdmal}). 
The potentials $V_{\rm b}$, $V_{\rm c}$ and $V_{\rm e}$ present some 
additional constant terms which do not arise from the area law. 
Finally there are not $1/r$ corrections as in the SVM. Some of these 
discrepancies can be interpreted as due to a finite thickness of the flux 
tube in DQCD opposite to the infinitely thin flux tube in the MAL 
model \cite{dualin}. Therefore in the two models the flux tube will have a 
different moment of inertia and give slightly different contributions to the 
velocity dependent potential. It is possible that these discrepancies will 
disappear if including higher order cumulants contributions in the 
SVM predictions. Other differences between the predictions  
of the two methods could have origin from the very delicate interpolating 
procedure of the numerical solutions of the DQCD non-linear equations.  
The very soon available lattice results on the velocity dependent 
potentials \cite{Bali} will possibly clarify the situation. 

\section{DISCUSSION AND CONCLUSIONS}

Using the same gauge invariant and physically transparent approach 
to calculate the complete semirelativistic quark-antiquark 
interaction for three different models (MAL, SVM and DQCD) we have 
shown the following points.
\begin{itemize}
\item{} 
We have obtained the velocity dependent corrections
in the SVM model which are new and  present an interesting 
non-perturbative structure.
\item{}
We have demonstrated that the minimal area law model 
is exactly reproduced in both the spin dependent and the 
velocity dependent sector of the potential by the long-range  
behaviour of the stochastic vacuum model. From now we can 
consider the MAL model simply as the $r\to\infty$ limit of 
the SVM for heavy quarks. Moreover this limit realizes 
also the intuitive Buchm\"uller's picture of zero magnetic 
field in the flux tube comoving system.
\item{}
In the spin dependent sector of the potential, 
both the SVM and DQCD not only reproduce the long-range behaviour  
given by the area law, but also give $1/r$ corrections  
to $dV_1/dr$ and $dV_2/dr$. These corrections are equal 
in both models and very near to the absolute value of the 
constant term in the static potential (the SVM also supplies
for the explication of this fact). This perfect agreement is 
absolutely not trivial and seems to be very meaningful, since it arises  
from two very different models in a region of distances 
in which the physics cannot be described by the area law alone.
This is also remarkable to understand the kind of effective kernel
that would describe the non-perturbative bound-states of constituent 
quarks. For example, it seems now clear that the vanishing 
of the magnetic part, given by the field average of eq. (\ref{vmag}),   
in the non-perturbative region takes place only at the leading level 
in the long-range limit. Therefore, working in a Bethe--Salpeter context, 
there is no need to assume an effective pure convolution kernel 
which is a Lorentz scalar (a recent proposed Bethe--Salpeter 
kernel can be found in \cite{bsnoi}).
\item{} 
Velocity dependent contributions to the quark-antiquark potential 
are important. In fact the string behaviour of the 
non-perturbative interaction shows up when we consider 
the velocity dependent part of the potential \cite{Dubin,dualin} 
and this is also what the data require \cite{letnoi}. 
The derivation of the velocity dependent part using equation 
(\ref{potential}) and the SVM is completely gauge invariant 
and seems not to suffer from the problems connected 
with the strong reduction dependence of the potentials obtained from 
Bethe--Salpeter kernels. In this way we reproduce the area law 
results and give a lot of new $1/r$ and $1/r^2$ corrections, suppressed 
in the long-range behaviour. The velocity dependent structure 
which arises from the DQCD model differs slightly in the coefficients 
with respect to the area law behaviour. The main reason seems to be that 
the flux tube in DQCD has a finite thickness. It is possible that higher 
order cumulants can reabsorb this difference.
\item{}
The spin dependent potentials have first been evaluated on the lattice. 
The data in \cite{lattice} confirm the long-range behaviour 
given in (\ref{vsmal}) and contained also in (\ref{vsosvm}) and 
(\ref{v0dqcd})-(\ref{v2dqcd}). Recent data \cite{lattice3} show up the same 
long-range behaviour and do not yet allow to distinguish between 
parametrizations which differ at the next-to-leading order in the 
distance $r$.  However they contain more information about the short-range 
region of the interaction (typically below the correlation length of 
0.2 fm). Generally the data reproduce the perturbative 
results (which at the first order in $\alpha_{\rm s}$ can be read from
(\ref{vsmal}) putting $\sigma$ equal 0). The only exception is given 
by the short distance behaviour of $dV_1/dr$ which seems to be 
negative and proportional to $1/r^2$. This contradicts the order 
$\alpha^2_{\rm s}$ calculation of the $Q\bar{Q}$ potential 
(which contains the first non vanishing perturbative contribution 
to $dV_1/dr$) given for example by Pantaleone and Tye \cite{panta}. 
The reason of this discrepancy could be explained 
by higher order perturbative contributions or 
by some at the moment unknown short range non-perturbative contribution 
(in the language of the SVM this contribution could arise from the 
correlation function $D$; an investigation in this sense of the 
recent short-range data on $D$ given in the last reference quoted in 
\cite{lattice2} is going on). The problem is still open. 
Only recently some data on the velocity dependent potentials appeared  
\cite{Bali,lattice2}. Probably more accurate data will be available in 
the next months. These results seem to confirm the long-range 
behaviour contained in (\ref{vdmal}) ($\sigma$ dependent terms). 
More interesting is the case of the potential $\Delta V_{\rm a}$ 
which appears to be different from zero for $r\to\infty$ and show up 
a $1/r$ short-range behaviour. This behaviour has been recently 
explained in terms of SVM and DQCD \cite{BVa}. 
\end{itemize} 

In conclusion SVM and DQCD reproduce the flux tube  
distribution measured on the lattice and the general features coming 
from the area law. Both give analytical expressions for the 
Wilson loop (eqs. (\ref{svm}) and (\ref{weff})) which 
describe the evolving behaviour of $\langle W(\Gamma)\rangle$ 
from the short to the long distances (we note that this can 
be useful in many different applications, see e.g. \cite{hera}) and 
both give some predictions which go beyond the asymptotic behaviour. 
But not all predictions are equal in the two models in the intermediate 
distances region, in particular in the velocity dependent sector of 
the potential, but also in the spin-spin interaction. 
Therefore, new lattice data sensitive to such kind of corrections 
seem to be urgent. Finally, work is in progress 
in evaluating the correlation function $D$ and $D_1$ 
in the DQCD context and in producing an extensive phenomenological 
analysis of the contribution of the new obtained potentials to 
the heavy and heavy-light quark spectrum.

\section*{Acknowledgements}

We would like to thank M. Baker, A. Di Giacomo, H. G. Dosch, D. Gromes, 
G. M. Prosperi, Yu.A. Simonov for enlightening conversations.
We also warmly acknowledge the kind hospitality given by the members
of the Theoretical Physics Institut of Heidelberg where part of this 
work was done.

\appendix
\section*{}

In this appendix we derive the static potential in the SVM 
(eq. (\ref{v0svm})). The same technique was used to obtain 
the other potentials. Since the velocity dependent potentials 
involve long and tedious calculations, a program of symbolic 
manipulations was used in that case \cite{Form}.

From eqs. (\ref{svm}), (\ref{cum2}) and in the straight-line 
parametrization of the surface, it follows that:
\begin{eqnarray}
\log \langle W (\Gamma) \rangle  &=& -{\beta\over 2} 
\int_S dS_{4i}(u) \int_S dS_{4j}(v) \Bigg[
\delta_{ij}\left( D\left( \tau^2 + ({\bf u} - {\bf v})^2  \right) 
+ D_1 \left(\tau^2 + ({\bf u} - {\bf v})^2 \right)  \right) 
\nonumber\\
&~& \quad\quad\quad
+ (\tau^2\delta_{ij} + (u-v)_i(u-v)_j){d\over d \tau^2} 
D_1 \left( \tau^2 +  ({\bf u} - {\bf v})^2 \right) \Bigg]
+ O({\dot z}^2_1,{\dot z}^2_2)  \>,
\label{A1}
\end{eqnarray}
with $dS_{4i}(u) = dt_1 ds_1 r_i(t_1)$, $dS_{4j}(v) = dt_2 ds_2 r_j(t_2)$ 
and $\tau \equiv t_1-t_2$. Expanding the functions of $t_2$ around 
$t_1$:
\begin{eqnarray}
r_j(t_2) &=& r_j(t_1) - {\dot r}_j(t_1) \tau + \dots \>,
\nonumber\\
(u-v)_j &=& z_{2j}(t_1) - z_{2j}(t_2) + s_1 r_j(t_1) - s_2 r_j(t_2) 
= (s_1-s_2)r_j(t_1) + \dots \>,
\nonumber 
\end{eqnarray}
and taking for simplicity $r_j(t_1) \equiv r_j$ and 
$\lambda \equiv s_2 - s_1$, we obtain  
\begin{eqnarray}
\log \langle W (\Gamma) \rangle  &=& -{\beta\over 2} 
\int_{t_{\rm i}}^{t_{\rm f}} dt_1
\int_{t_{\rm i}}^{t_{\rm f}} dt_2
\int_0^1 ds_1 \int_0^1 ds_2 ~r_ir_j \Bigg[
\delta_{ij}\left( D(\tau^2 + \lambda^2 r^2) + D_1(\tau^2 + \lambda^2 r^2) 
\right)  
\nonumber\\
&~& \quad\quad\quad\quad\quad
+ (\tau^2\delta_{ij} + \lambda^2 r_i r_j){d\over d \tau^2} 
D_1(\tau^2 + \lambda^2 r^2) \Bigg]
+ O({\dot z}^2_1,{\dot z}^2_2)  \>.
\label{A2}
\end{eqnarray}
Since:
$$
\int_0^1 ds_1 \int_0^1 ds_2 ~f\left( (s_2-s_1)^2 \right)  = 
\int_0^1 ds_1 \int_{-s_1}^{1-s_1} d\lambda~f(\lambda^2) = 
2\int_0^1 d\lambda~(1-\lambda)~ f(\lambda^2) \>,
$$
we can write
\begin{eqnarray}
\log \langle W (\Gamma) \rangle  &=& -\beta 
\int_{t_{\rm i}}^{t_{\rm f}} dt_1
\int_{t_{\rm i}}^{t_{\rm f}} dt_2
\int_0^1 d\lambda ~(1-\lambda)~ r^2 \Bigg[
D(\tau^2 + \lambda^2 r^2) + D_1(\tau^2 + \lambda^2 r^2)  
\nonumber\\
&~& \quad\quad\quad\quad\quad
+ (\tau^2 + \lambda^2 r^2){d\over d \tau^2} 
D_1(\tau^2 + \lambda^2 r^2) \Bigg] + O({\dot z}^2_1,{\dot z}^2_2)  \>.
\label{A3}
\end{eqnarray}
Replacing $r \lambda \to \lambda$  and taking in account that the 
time variables in (\ref{A3}) are in an Euclidian space while the equation  
for the potential (\ref{potential}) is in Minkowski, the static potential 
is given by
\begin{eqnarray}
V_0(r) &=& \beta \int_{-\infty}^{+\infty} d\tau \int_0^r d\lambda ~
(r-\lambda)~ D(\tau^2 + \lambda^2)
\nonumber\\
&+& \beta \int_{-\infty}^{+\infty} d\tau \int_0^r d\lambda ~
(r-\lambda)~ \left[ D_1(\tau^2 + \lambda^2) + 
(\tau^2 + \lambda^2){d\over d \tau^2} D_1(\tau^2 + \lambda^2) \right] \>,
\label{A4}
\end{eqnarray}
where, also, the large time limit was performed. 
Finally, the identities
$$
\int_{-\infty}^{+\infty} d\tau ~\tau^2 {d\over d \tau^2} 
D_1(\tau^2 + \lambda^2)  = 
{1\over 2} \int_{-\infty}^{+\infty} d\tau ~\tau {d\over d \tau} 
D_1(\tau^2 + \lambda^2)  = 
- {1\over 2} \int_{-\infty}^{+\infty} d\tau ~ D_1(\tau^2 + \lambda^2) \>,
$$
and 
\begin{eqnarray}
\int_0^r d\lambda ~(r-\lambda) ~\lambda^2  {d\over d \tau^2} 
D_1(\tau^2 + \lambda^2)  &=& 
{1\over 2} \int_0^r d\lambda ~(r-\lambda) ~\lambda  {d\over d \lambda} 
D_1(\tau^2 + \lambda^2)  
\nonumber\\
&~& \quad\quad 
= - {1\over 2} \int_0^r d\lambda ~(r- 2~\lambda) ~ D_1(\tau^2 + \lambda^2) 
\>,
\nonumber
\end{eqnarray}
give back the static potential in the form of eq. (\ref{v0svm}). 
Taking in account the $O({\dot z}^2_1,{\dot z}^2_2)$ contributions 
in (\ref{A1}) and in the following equations, we obtain the velocity 
dependent potentials (\ref{vbsvm})-(\ref{vesvm}).

\vfill
\eject

\begin{figure}[htb]
\vskip 0.8truecm
\makebox[0.2truecm]{\phantom b}
\epsfxsize=14.8truecm
\epsffile{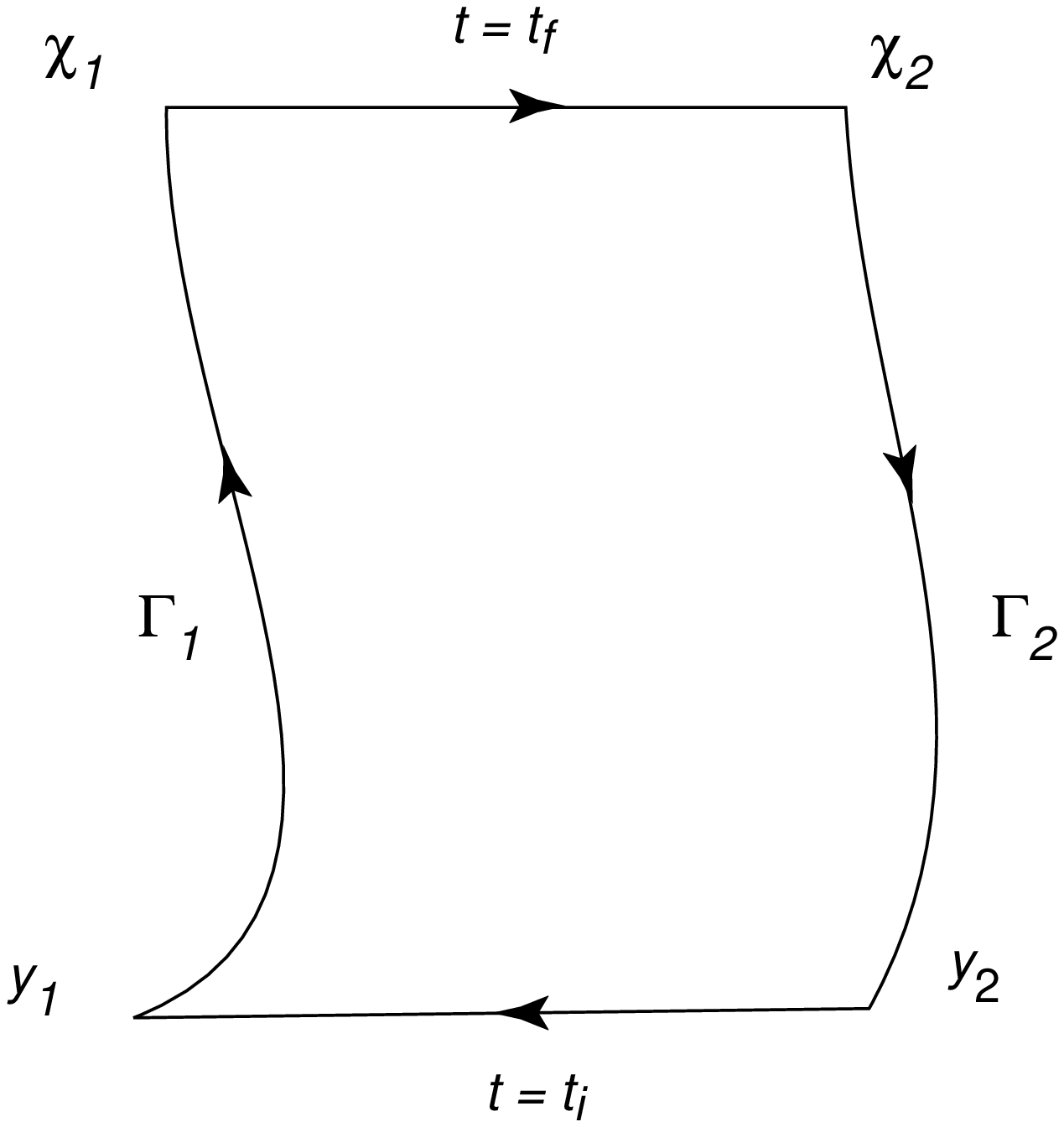}
\vskip 0.3truecm
\caption{Quark-antiquark Wilson loop}
\label{qqloop}
\vskip 0.8truecm
\end{figure}


\begin{references}
\bibitem{Wilson} K. G. Wilson, {Phys. Rev.} {\bf D 10}, {2445} (1974);  
\bibitem{Kummer} W. Kummer and W. M\"odritsch, Z. Phys. {\bf C 66}, 225   
 (1995); W. Kummer, W. M\"odritsch and A. Vairo,  
 CERN-TH/96-27, hep-ph/9602276 (1996) (Zeit. Phys. {\bf C} in press);  
\bibitem{Buch} W. Buchm\"uller, {Phys. Lett.} {\bf B 112}, 479 (1982);
\bibitem{eichten} E. Eichten and F. Feinberg, {Phys. Rev.} {\bf D 23}, 2724 
 (1981); D. Gromes, in Proceedings of the International School of 
 Physics ``Ettore Majorana", International Science Series 
 {Vol. {\bf 37}} 67 (Plenum, New York, 1989); 
 M. A. Peskin, in Proceedings of the 11th SLAC Inst., 
 SLAC Rep. n.207, 151  ed. by P. Mc. Donough (1993);
\bibitem{Gromes} D. Gromes, {Z. Phys.} {\bf C 26}, 401  (1984);
\bibitem{lattice} M. Campostrini, K. Moriarty and C. Rebbi, Phys. Rev. Lett.
 {\bf 57}, 44 (1986); K.D. Born, E. Laermann, R. Sommer, T.F. Walsh and
 P.M. Zerwas, Phys. Lett. {\bf B 329} 325 (1994); 332 (1994); 
\bibitem{lattice3} G. S. Bali, K. Schilling and A. Wachter, Nucl. Phys. 
 {\bf B} Proc. Suppl. 42, 213 (1995); in Proceedings of ``Confinement '95'' 
 eds. H. Toki et al., p. 82, (World Scientific, Singapore, 1995); 
\bibitem{HQET} Yu-Qi Chen, Yu-Ping Kuang and R. J. Oakes,
 Phys. Rev. {\bf D52} 264 (1995);
\bibitem{Gupta} S. N. Gupta, S. F. Randford and W. W. Repko,  
 Phys. Rev. {\bf D 34}, 201 (1986);
\bibitem{Durand} A. Gara, B. Durand and L. Durand, Phys. Rev.
 {\bf D 42}, 1651 (1990); {\bf D 40}, 843 (1989);
\bibitem{Lagae} J. F. Lagae, Phys. Rev. {\bf D 45}, 305 (1992); 
  317 (1992); N. Brambilla and G. M. Prosperi, Phys. Rev.
 {\bf D 48}, 2360 (1993); {\bf D 46}, 1096 (1992);
\bibitem{BCP} N. Brambilla, P. Consoli and G. M. Prosperi, Phys. Rev.
 {\bf D 50},  5878 (1994); N. Brambilla and G. M. Prosperi, in
 Proceedings of ``Quark Confinement and the Hadron Spectrum'', eds. 
 N. Brambilla and G. M. Prosperi, p. 113, (World Scientific,   
 Singapore, 1995);
\bibitem{BMP} A. Barchielli, E. Montaldi and G. M. Prosperi,
 Nucl. Phys.  {\bf B 296}, {625} (1988);
\bibitem{NC} A. Barchielli, N. Brambilla and G. M. Prosperi,
 Nuovo Cimento {\bf 103 A}, 59 (1990);
\bibitem{Sinp} Yu. A. Simonov, {Nucl. Phys.} {\bf B 307}, 512 (1988);
 {\bf B 324}, 67 (1989);
\bibitem{Dubin} A. Yu. Dubin, A. B. Kaidalov and Yu. A. Simonov, 
 Phys. Lett. {\bf B 323}, 41 (1994);
\bibitem{Bali} G. Bali, private communications, see also the
 contribution of G. Bali in Proceedings of ``Quark Confinement
 and the Hadron spectrum II", eds. N. Brambilla
 and G. M. Prosperi (World Scientific, Singapore);
\bibitem{dual} M. Baker, J. Ball and F. Zachariasen, Phys. Rev.
 {\bf D 51}, 1968 (1995); M. Baker, J. S. Ball and F.
 Zachariasen,  Phys. Rep. {\bf 209}, 73 (1991); M. Baker, J. S. Ball
 and F. Zachariasen, Phys. Lett. {\bf B 283}, 360 (1992);
\bibitem{dualin} M. Baker, J. Ball, N. Brambilla, G. M. Prosperi and  
 F. Zachariasen, Phys. Rev. {\bf D 54}, 2829 (1996);
\bibitem{ols2} D. La Course and M. G. Olsson, {Phys. Rev.} {\bf D 39}, 2751
 (1989) and references therein; M. G. Olsson in Proceedings 
 of ``Quark Confinement and the Hadron spectrum", eds. N. Brambilla 
 and G. M. Prosperi, p.76 (World Scientific, Singapore, 1995); 
 M. G. Olsson, and S. Veseli, Phys. Rev {\bf D 53}, 4006 (1996);
\bibitem{fluxnoi} N. Brambilla and G. M. Prosperi, Phys. Rev. 
 {\bf D 47}, 2107 (1993); N. Brambilla, G. M. Prosperi and 
 A. Vairo, Phys. Lett. {\bf B 362}, 113 (1995);
\bibitem{report} W. Lucha, F. F. Sch\"oberl and D. Gromes, 
 {Phys. Rep.} {\bf 200}, 127 (1990);
\bibitem{letnoi} N. Brambilla and G. M. Prosperi, Phys. Lett.
 {\bf B 236}, 69 (1990);
\bibitem{Olsson} M. G. Olsson, S. Veseli and K. Williams, 
 Phys. Rev. {\bf D 52}, 5141 (1995); {\bf D 53}, 504 (1996);
\bibitem{DoSi} H. G. Dosch, Phys. Lett. {\bf B 190}, 177 (1987); 
 H. G. Dosch and Yu. A. Simonov, Phys. Lett. {\bf B 205}, 339 (1988); 
 M. Schiestl and H. G. Dosch, Phys. Lett. {\bf B 209}, 85 (1988); 
\bibitem{DoSi2} Yu. A. Simonov, Yad. Fiz. {\bf 54}, 192 (1991);
 H. G. Dosch, Prog. Part. Nucl. Phys. {\bf 33}, 121 (1994);
\bibitem{rueter} M. Rueter, H. G. Dosch, Z. Phys. {\bf C 66}, 245 (1995); 
\bibitem{back} Yu. A. Simonov Phys. At. Nucl. {\bf 58}, 107 (1995); 
 in Proceedings of ``Perturbative and Nonperturbative Aspects 
 of Quantum Field Theory'', (Schladming, 1996);
\bibitem{Stokes} I. Ya. Aref'eva, Teor. Mat. Fiz. {\bf 43}, 
 111 (1980); N. Bralic, Phys. Rev. {\bf D 22}, 3090 (1980); 
 P. M. Fishbane, S. Gasiorowicz and P. Kaus, Phys. Rev. 
 {\bf D 24}, 2324 (1981); 
\bibitem{Si} Yu. A. Simonov Yad. Fiz. {\bf 50}, 213 (1989);
\bibitem{Kampen} N. G. Van Kampen, Phys. Rep. {\bf 24 C}, 171 (1976);
\bibitem{lattice2} M. Campostrini, A. Di Giacomo and G. Mussardo, 
 Z. Phys. {\bf C 25}, 173 (1984); A. Di Giacomo and H. Panagopoulos 
 Phys. Lett. {\bf B 285}, 133 (1992); A. Di Giacomo, E. Meggiolaro
 and H. Panagopoulos, (March 1996) hep-lat/9603017; 
\bibitem{BaSi} A. M. Badalian and Yu. A. Simonov, 
 {\it Spin-dependent potentials and field correlators in QCD} 
 in preparation (1996);
\bibitem{BaYu} A. M. Badalian and V. P. Yurov, Yad. Fiz. {\bf 51}, 1368 
 (1990); Phys. Rev. {\bf D 42}, 3138 (1990); 
\bibitem{BVa}  M. Baker, J. S. Ball, N. Brambilla and A. Vairo, {\em 
 Nonperturbative evaluation of a field correlator appearing in the 
 heavy quarkonium system}, IFUM 537/FT, hep-ph/9609233 (1996) 
 (Phys. Lett. {\bf B} in press); 
\bibitem{Mandelstam} S. Mandelstam, Phys. Rep. {\bf 23 C}, 145 (1976);
 G. t'Hooft, in "Proc. Eur. Phys. Soc. 1975", 1225, ed. by A. Zichichi
 (Ed. Comp. Bologna 1976);
\bibitem{Dirac}P. A. M. Dirac, Phil. Mag. {\bf 39}, 537 (1920);
\bibitem{baker} M. Baker in Proceedings of the ``Workshop 
 on Quantum Infrared Physics", Eds. H.M. Fried, B. Muller, 
 (World Scientific, Singapore, 1995); M. Baker, J. S. Ball and F. Zachariasen, 
 Int. Jour. Mod. Phys. {\bf A 11}, 343 (1996); 
 A. M. Green, C. Michael and P. S. Spencer, HU-TFT-96-36 in Proceedings 
 of ``Quark Confinement and the Hadron spectrum II", eds. N. Brambilla
 and G. M. Prosperi (World Scientific, Singapore);
\bibitem{bsnoi} N. Brambilla, E. Montaldi and G. M. Prosperi,
 Phys. Rev. {\bf D 54}, 3506 (1996); 
\bibitem{panta} J. Pantaleone and S. H. Tye, Phys. Rev. {\bf D 37}, 
 3337 (1988); 
\bibitem{hera}H. G. Dosch, E. Ferreira, A. Kr\"amer, Phys. Rev.
 {\bf D 50}, 1992 (1994);
\bibitem{Form}J. A. M. Vermaseren, {\it Symbolic manipulation with FORM}, 
 (Computer Algebra Nederland, Amsterdam, 1991). 
\end{references}
\end{document}